\documentclass[10pt]{amsart}
\usepackage{amssymb}
\usepackage{enumerate}
\usepackage[dvips]{graphicx}
\DeclareGraphicsExtensions{.eps,.art,.ART,.ps}
\newcommand{\bbR}{\mathbb{R}}      % real numbers %
\newtheorem{Theo}{Theorem}[section]
\newtheorem{Prop}[Theo]{Proposition}

\newtheorem{Cor}[Theo]{Corollary}
\theoremstyle{definition}
\newtheorem{Def}[Theo]{Definition}
\theoremstyle{remark}
\newtheorem{exmp}[Theo]{Example}

\begin{document}
\begin{abstract}
It is commonly believed that Alcubierre's warp drive works by contracting space in front of the warp bubble and expanding space behind it. We show that this contraction/expansion is but a marginal consequence of the choice made by Alcubierre, and explicitly construct a similar spacetime where no contraction/expansion occurs. Global and optical properties of warp drive spacetimes are also discussed.  
\end{abstract}
%
%%%%%%%%%%%%%%%%%%%%%%%%%%%% Definition of title page %%%%%%%%%%%%%%%%%%%%%%%%%%%
%
\title{Warp Drive With Zero Expansion}
\author{Jos\'{e} Nat\'{a}rio}
\address{Department of Mathematics, Instituto Superior T\'{e}cnico, Portugal}
%\subjclass[2000]{Primary 53078}
\thanks{This work was partially supported by FCT (Portugal) through programs PRAXIS XXI and PROCTI}
\maketitle
%
%
%%%%%%%%%%%%%%%%%%%%%%%%%%%%%%%%%%% Section 1 %%%%%%%%%%%%%%%%%%%%%%%%%%%%%%%%%%
%
\section*{Introduction}
Alcubierre's warp drive spacetime (\cite{A94}) allows effective superluminal travel within General Relativity. As remarked by Alcubierre himself, it requires negative energy densities, and there are reasons to believe that all spacetimes allowing superluminal travel must violate energy conditions (\cite{VBL00}, \cite{GW00}). Another feature is the existence of horizons (\cite{CHL99}), which has been argued on general grounds to be unavoidable (\cite{L99}).

The traditional heuristic explanation of how the warp drive spacetime works is that space in front of a given region (the ``warp bubble'') is contracted, whereas space behind the same region is expanded. In response, the warp bubble moves forwards with a speed determined by the contraction/expansion rate.

We will show in this paper that this contraction/expansion is not necessary at all, and that in particular it is possible to construct a similar spacetime where no contraction/expansion occurs. Heuristically, one could best describe the warp drive spacetime as ``sliding'' the warp bubble region through space; space in front of the bubble may get contracted (and space behind it expanded), or not, depending on the details of the construction.

The paper is divided in three sections. 

In the first section we define the warp drive spacetime generated by any (time-dependent) vector field ${\bf X}$ in Euclidean 3-space. This is a globally hyperbolic spacetime foliated by Euclidean 3-spaces such that the integral lines of {\bf X} represent timelike geodesics (corresponding to the so-called {\em Eulerian observers}). We show that such spacetimes always violate some energy condition, and compute the expansion of the Eulerian observers' volume element to be $\nabla \cdot {\bf X}$. Finally, we show how Alcubierre's warp drive can be obtained from a particular choice of ${\bf X}$. 

In the second section, we explicitly construct a volume-preserving warp drive by selecting an appropriate divergenceless vector field ${\bf X}$.

In the third section, we take advantage of the formalism developed thus far to analyse some simple examples, and find them to possess horizons and infinite blueshifts. These results are similar to those obtained (some numerically) in \cite{CHL99}, but independent of the particular choice of warp drive, and perhaps easier to visualize.
%
%
%%%%%%%%%%%%%%%%%%%%%%%%%%%%%%%%%%% Section 1 %%%%%%%%%%%%%%%%%%%%%%%%%%%%%%%%%%
%
\section{Warp Drive Spacetimes}
\begin{Def}
A {\em warp drive spacetime} is a globally hyperbolic spacetime $(M,g)$, where $M=\bbR^4$ with the usual Cartesian coordinates $(t,x,y,z)\equiv (t,x^i)$ and $g$ is given by the line element\footnote{We shall use the notation and sign conventions in \cite{W84}, except for the use the latin indices $i,j,k$ as {\em numerical} indices running from 1 to 3 and referring either to the space coordinates $(x,y,z)$ or to space components in an orthonormal frame including $\left\{\frac{\partial}{\partial x},\frac{\partial}{\partial y},\frac{\partial}{\partial z}\right\}$.}
\[
ds^2=-dt^2+\sum_{i=1}^3(dx^i-X^idt)^2
\]
for three unspecified bounded smooth functions $(X^i)\equiv (X,Y,Z)$.
\end{Def}
\noindent Notice that the Riemannian metric $h$ induced in the Cauchy surfaces $\{dt=0\}$ is just the ordinary Euclidean flat metric. A warp drive spacetime is completely defined by the vector field
\[
{\bf X}=X^i\frac{\partial}{\partial x^i}=X\frac{\partial}{\partial x}+Y\frac{\partial}{\partial y}+Z\frac{\partial}{\partial z},
\]
which we can think of as a (time-dependent) vector field defined in Euclidean 3-space. The future-pointing unit normal covector to the Cauchy surfaces is
\[
n_a=-dt \Leftrightarrow n^a=\frac{\partial}{\partial t}+X^i\frac{\partial}{\partial x^i}=\frac{\partial}{\partial t}+{\bf X}.
\]
\begin{Def}
The observers whose 4-velocity is $n^a$ are said to be the {\em Eulerian observers}.
\end{Def}
\begin{Prop}
Eulerian observers are free-fall observers.
\end{Prop}
\begin{proof}
Using the geodesic Lagrangian,
\[
L=\frac12\left[-\dot{t}^2+\sum_{i=1}^3\left(\dot{x}^i-X^i\dot{t}\right)^2\right],
\]
it is a simple matter to write down the geodesic equations:
\begin{align*}
& \frac{d}{d\tau}\left(\frac{\partial L}{\partial \dot{t}}\right)-\frac{\partial L}{\partial t}=0 \Leftrightarrow \frac{d}{d\tau}\left(-\dot{t}-X^i\left(\dot{x}^i-X^i\dot{t}\right)\right)+\dot{t}\frac{\partial X^i}{\partial t}\left(\dot{x}^i-X^i\dot{t}\right)=0; \\
& \frac{d}{d\tau}\left(\frac{\partial L}{\partial \dot{x}^i}\right)-\frac{\partial L}{\partial x^i}=0 \Leftrightarrow \frac{d}{d\tau}\left(\dot{x}^i-X^i\dot{t}\right)+\dot{t}\frac{\partial X^j}{\partial x^i}\left(\dot{x}^j-X^j\dot{t}\right)=0.
\end{align*}
It is immediate to check that any curve satisfying $\dot{t}=1, \dot{x}^i = X^i$ is a solution of these equations.
\end{proof}
\begin{Prop}
The extrinsic curvature tensor associated to the foliation of a warp drive spacetime by the Cauchy surfaces $\{dt=0\}$ is
\[
K =\frac12\left(\partial_iX^j + \partial_jX^i \right)dx^i \otimes dx^j.
\]
\end{Prop}
\begin{proof}
As is well known, the extrinsic curvature tensor is given by
\[
K = \frac12 \pounds_{n^a} h = \frac12 \pounds_{\frac{\partial}{\partial t}+{\bf X}} h. 
\]
Now
\[
\pounds_{\frac{\partial}{\partial t}}h = \pounds_{\frac{\partial}{\partial t}}\delta_{ij}dx^i \otimes dx^j = \frac{\partial \delta_{ij}}{\partial t}dx^i\otimes dx^j=0.
\]
On the other hand, since ${\bf X}$ is tangent to the Cauchy surfaces, we can use the usual formula for the Lie derivative of the metric,
\[
\pounds_{\bf X}h = \left( \delta_{kj} D_iX^k+\delta_{ik}D_jX^k\right)dx^i \otimes dx^j = \left( D_iX^j+D_jX^i\right)dx^i \otimes dx^j,
\]
where $D$ stands for the Levi-Civita connection determined by $h$. Since $h$ is just the flat Euclidean metric, $D=\partial$ and we get the formula above.
\end{proof}
\begin{Cor}
The expansion of the volume element associated with the Eulerian observers is $\theta = \nabla \cdot {\bf X}$.
\end{Cor}
\begin{proof}
We just have to notice that
\[
\theta = K^i_{\,\,i}=\partial_iX^i.
\]
\end{proof}
\begin{Cor}
A warp drive spacetime is flat wherever $\bf X$ is a Killing vector field for the Euclidean metric (irrespective of time dependence). In particular, a warp drive spacetime is flat wherever $\bf X$ is spatially constant.
\end{Cor}
\begin{proof}
Since the Cauchy surfaces are flat, all curvature comes from the extrinsic curvature. Thus the spacetime will be flat wherever the extrinsic curvature is zero, {\em i.e.}, wherever $\pounds_{\bf X}h=0$.
\end{proof}
\begin{Theo}
Non flat warp drive spacetimes violate either the weak or the strong energy condition.
\end{Theo}
\begin{proof}
If the strong energy condition holds and $\theta \neq 0$ at some event, then $\theta$ blows up in finite time (see \cite{LL97}). Since $\nabla \cdot {\bf X}$ is finite, the strong energy condition can only hold if $\theta \equiv 0$. However, it follows from Einstein's field equations that the energy density measured by Eulerian observers is (see \cite{W84})
\[
\rho=T_{ab}n^an^b=\frac{1}{16 \pi}\left( ^{(3)}R+(K^i_{\,\,i})^2 - K_{ij}K^{ij}\right)=\frac{1}{16 \pi}\left( \theta^2 - K_{ij}K^{ij}\right)
\]
where $^{(3)}R=0$ is the scalar curvature of the flat Cauchy surfaces $dt=0$. Thus if $\theta=0$ we have $\rho\leq 0$, and $\rho=0$ {\em iff} $K_{ij}\equiv 0$. Consequently if the spacetime does not violate neither the strong nor the weak energy conditions it must be flat.
\end{proof}
\begin{exmp}
Alcubierre's warp drive (see \cite{A94}) is obtained by choosing
\begin{align*}
& X=v_s f(r_s); \\
& Y=Z=0
\end{align*}
with
\begin{align*}
& v_s(t)=\frac{dx_s(t)}{dt}; \\
& r_s=\left[ (x-x_s)^2+y^2+z^2 \right]^\frac12,
\end{align*}
for a smooth function $f:[0, +\infty)\to[0,1]$ approximating a step function equal to $1$ in a neighbourhood of the origin and equal to $0$ in a neighbourhood of infinity, and an arbitrary function $x_s(t)$. If we call the region $f(r_s)=1$ the {\em interior of the warp bubble} and the region $f(r_s)=0$ the {\em exterior of the warp bubble}, we see that both these regions are flat; nevertheless, Eulerian observers inside the warp bubble move with arbitrary speed $v_s$ with respect to Eulerian observers outside the warp bubble.
It is a simple matter to obtain the expansion of the volume element associated with the Eulerian observers and the  energy density measured by Eulerian observers in this example:
\begin{align*}
& \theta = \partial_x X = v_s f'(r_s) \frac{x-x_s}{r_s}; \\
& \rho = \frac{1}{16 \pi} \left[ \left(\partial_xX\right)^2 - \left(\partial_xX\right)^2 - 2\left(\frac12 \partial_y X\right)^2 - 2\left(\frac12 \partial_z X\right)^2 \right]  \\
& \,\,\,\, = -\frac{1}{32\pi}v_s^2 \left[ f'(r_s)\right]^2 \frac{y^2+z^2}{r_s^2}.
\end{align*}
It is convenient to replace the $x$ coordinate with
\[
\xi = x - x_s(t).
\]
This effectively corresponds to replacing $X$ with $X-v_s$, so that the Eulerian observers inside the bubble stand still whereas the Eulerian observers outside the bubble move with speed $v_s$ in the negative $\xi$-direction. Obviously $\theta$ and $\rho$ retain their values, but now
\[
r_s=\left(\xi^2+y^2+z^2\right)^\frac12
\]
does not depend on the coordinate $t$.
\end{exmp}

\begin{Def} The vector field {\bf X} is said to {\em generate a warp bubble with velocity} ${\bf v}_s(t)$ if  ${\bf X}={\bf 0}$ for small $\|{\bf x}\|$ (the {\em interior of the warp bubble}) and ${\bf X}=-{\bf v}_s(t)$ for large  $\|{\bf x}\|$ (the {\em exterior of the warp bubble}), where
\[
{\bf x}=x^i\frac{\partial}{\partial x^i}=x\frac{\partial}{\partial x}+y\frac{\partial}{\partial y}+z\frac{\partial}{\partial z}.
\]
\end{Def}
%
%
%%%%%%%%%%%%%%%%%%%%%%%%%%%%%%%%%%%%%%% Section 2 %%%%%%%%%%%%%%%%%%%%%%%%%%%%
%
\section{Warp Drive With Zero Expansion}
We introduce spherical coordinates $(r, \theta, \varphi)$ in the Euclidean 3-space with the $x$ axis as the polar axis. Using the usual isomorphism $\bbR^3 \sim \Lambda^1\bbR^3 \sim \Lambda^2\bbR^3$ provided by the Euclidean metric and the Hodge star, given in spherical coordinates by
\begin{align*}
& {\bf e}_r \equiv \frac{\partial}{\partial r} \sim dr \sim (rd\theta) \wedge (r \sin \theta d \varphi); \\
& {\bf e}_\theta \equiv \frac1r \frac{\partial}{\partial \theta} \sim rd\theta \sim (r \sin \theta d\varphi) \wedge dr; \\
& {\bf e}_\varphi \equiv \frac1{r \sin \theta} \frac{\partial}{\partial \varphi} \sim r \sin \theta d\varphi \sim dr \wedge (rd\theta),
\end{align*}
we have
\begin{align*}
& \frac{\partial}{\partial x} \sim dx = d(r\cos\theta) = \cos \theta dr - r \sin \theta d\theta \sim \\
& \sim r^2 \sin \theta \cos \theta d\theta \wedge d\varphi + r \sin^2 \theta dr \wedge d\varphi = d \left( \frac12 r^2 \sin^2 \theta d\varphi \right).
\end{align*}
Consequently if we set
\begin{align*}
{\bf X} \sim -v_s(t)  d \left[ f(r)  r^2 \sin^2 \theta d\varphi \right] \sim -2v_s f\cos\theta {\bf e}_r+v_s(2f+rf')\sin \theta {\bf e}_\theta
\end{align*}
with $f(r)=\frac12$ for large $r$ and $f(r)=0$ for small $r$, we are guaranteed to have a divergenceless field generating a warp bubble with velocity  $v_s(t)\frac{\partial}{\partial x}$. Interestingly, formulae for the extrinsic curvature tensor
\[
K_{ij}=\frac12 \left(\partial_iX^j + \partial_jX^i\right)
\]
in spherical coordinates are available, as this corresponds to the so-called {\em rate-of-strain} tensor of fluid mechanics (see for instance \cite{A98}). If we set
\[
{\bf X}=X^r {\bf e}_r +X^\theta {\bf e}_\theta +X^\varphi {\bf e}_\varphi 
\]
we have
\begin{align*}
& K_{rr}=\frac{\partial X^r}{\partial r}=-2v_sf'\cos\theta; \\
& K_{\theta \theta}=\frac1r\frac{\partial X^\theta}{\partial \theta}+\frac{X^r}{r}=v_sf'\cos\theta; \\
& K_{\varphi \varphi}=\frac1{r \sin \theta}\frac{\partial X^\varphi}{\partial \varphi}+\frac{X^r}{r}+\frac{X^\theta \cot \theta}{r}=v_sf'\cos\theta; \\
& K_{r\theta}=\frac12 \left[ r\frac{\partial}{\partial r}\left(\frac{X^\theta}{r}\right)+\frac1r\frac{\partial X^r}{\partial \theta}\right]=v_s\sin\theta\left(f'+\frac{r}2f''\right); \\
& K_{r\varphi}=\frac12 \left[ r\frac{\partial}{\partial r}\left(\frac{X^\varphi}{r}\right)+\frac1{r\sin\theta}\frac{\partial X^r}{\partial \varphi}\right]=0; \\
& K_{\theta\varphi}=\frac12 \left[ \frac{\sin\theta}{r}\frac{\partial}{\partial \theta}\left(\frac{X^\varphi}{\sin \theta}\right)+\frac1{r\sin\theta}\frac{\partial X^\theta}{\partial \varphi}\right]=0,
\end{align*}
yielding the energy density (as measured by Eulerian observers)
\[
\rho = -\frac{1}{16\pi}K_{ij}K^{ij}=-\frac{v_s^2}{8\pi}\left[ 3(f')^2 \cos^2 \theta +  \left(f' + \frac{r}2 f''\right)^2\sin^2\theta\right].
\]
Notice that
\[
\theta = K_{rr}+K_{\theta\theta}+K_{\varphi\varphi}=0,
\]
as it should. This gives some insight into the geometry of this spacetime: for example for $\cos \theta>0$ (i.e., in front of the bubble), $K_{rr}<0$ where $f'>0$ (i.e., in the bubble's wall), indicating a compression in the radial direction; this is however exactly balanced by $K_{\theta \theta} + K_{\varphi \varphi} = - K_{rr}$, corresponding to an expansion in the perpendicular direction.  
%
%
%%%%%%%%%%%%%%%%%%%%%%%%%%%%%%%%%%%%%%% Section 3 %%%%%%%%%%%%%%%%%%%%%%%%%%%%
%
\section{Null geodesics and horizons}
Besides violating energy conditions, warp drive spacetimes may have other pathologies, such as horizons and infinite blueshifts. To see evidence of this, let us consider the case of a vector field ${\bf X}$ generating a warp bubble with velocity  $v_s\frac{\partial}{\partial x}$ satisfying
\[
\frac{\partial{\bf X}}{\partial t}={\bf 0} \,\, \left(\Rightarrow \frac{dv_s}{dt}=0\right).
\]
Notice that
\[
< \frac{\partial}{\partial t} , \frac{\partial}{\partial t}> = -1+{\bf X}^2
\]
and hence such spacetime can only be stationary if $|v_s|<1$.

\begin{figure}[h]
\begin{center}    
        %\leavevmode
        \includegraphics[scale=.4]{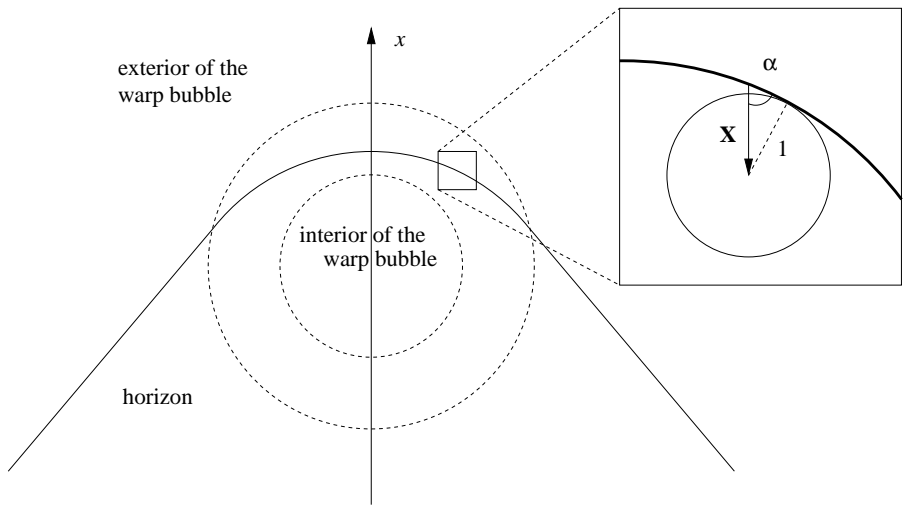}
\end{center}
\caption{Computing the horizon.}
\label{horizon}
\end{figure}

Let us take $v_s > 1$. Since null geodesics must satisfy
\[
ds^2 = 0 \Leftrightarrow dt^2=\sum_{i=1}^3(dx^i-X^idt)^2 \Leftrightarrow \left\|\frac{d{\bf x}}{dt}-{\bf X}\right\|=1
\]
we see that a flash of light outside the warp bubble can be pictured in the Euclidean 3-space as a spherical wavefront which is simultaneously expanding with speed 1 and moving in the direction of ${\bf X}$ with speed $\|{\bf X}\|=v_s$. Thus it is clear that events inside the warp bubble cannot causally influence events outside the warp bubble at large positive values of $x$. Assuming cylindrical symmetry about the $x$-axis, there will be a point on the positive $x$-axis where $\|{\bf X}\|=1$; the cylindrically symmetric surface through this point whose angle $\alpha$ with ${\bf X}$ is given by
\[
\sin \alpha = \frac{1}{\|{\bf X}\|}
\]
is a horizon, in the sense that events inside the warp bubble cannot causally influence events on the other side of this surface (see figure \ref{horizon}). Notice that away from the warp bubble we have
\[
\sin \alpha =\frac{1}{v_s}
\]
which is the familiar expression for the Mach cone angle. Notice also that the interior of the warp bubble is causally disconnected from part of the bubble's wall, as is unavoidable (see \cite{L99}).

\begin{figure}[h]
\begin{center}    
        %\leavevmode
        \includegraphics[scale=.4]{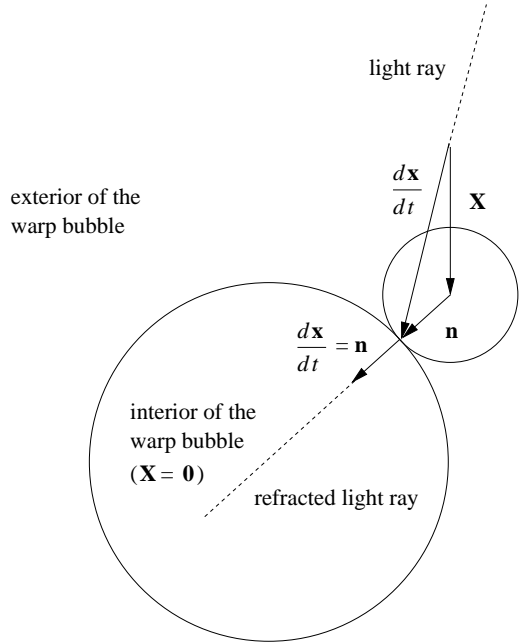}
\end{center}
\caption{Refraction at the warp bubble's wall.}
\label{horizon2}
\end{figure}

Inside and outside the warp bubble, the warp drive spacetime is flat, and hence the null geodesics are straight lines. If we assume that the warp bubble's wall is very thin, to study light rays crossing it all we have to decide is how much these light rays are refracted. According to the {\em Fermat principle} of General Relativity (see \cite{SEF92}), and since the proper time of an Eulerian observer inside the bubble is just $t$, we must minimize the $t$ coordinate of the event at which the observer sees the light ray. If we choose the observer sitting at the centre of the bubble, light from every point in the bubble's wall will reach him simultaneously; thus this observer will see the first light ray emitted at a given event to reach the warp bubble's wall. If ${\bf x}(t)$ is the position of the light ray at time coordinate $t$, the unit vector
\[
{\bf n} = \frac{d{\bf x}}{dt}-{\bf X}
\]
is orthogonal to the wavefront (and in fact represents the direction of the corresponding light ray from the point of view of a Eulerian observer outside the warp bubble). The first light ray to reach the warp bubble will do so at a tangency between the wavefront and the warp bubble. Therefore ${\bf n}$ must be orthogonal to the warp bubble, and will therefore coincide with the direction of the light ray {\em inside} the warp bubble (see figure \ref{horizon2}). Incidentally, there will be no aberration: the observer in the centre of the bubble will see light coming from the same direction as seen by an Eulerian observer outside the bubble in the light ray's path\footnote{This however is slightly misleading: it is not hard to show that the expression for the refraction angle $r$ (measured by an Eulerian observer inside the bubble) in terms of the incidence angle $i$ (measured by an Eulerian observer outside the bubble) at a point in the bubble's wall making an angle $\theta$ with the $x$-axis is
\[
\sin r = \frac{\sin i}{1+v_s \cos(i-\theta)}.
\]
Thus by making $v_s$ large enough there will be light coming from arbitrary directions arbitrarily close to the centre of the bubble.}. It is possible to show that this is the limit of the results obtained in \cite{CHL99} as the thickness of bubble's wall tends to zero. If the thickness is finite, however, there will always be some aberration.

\begin{figure}[h]
\begin{center}    
        %\leavevmode
        \includegraphics[scale=.4]{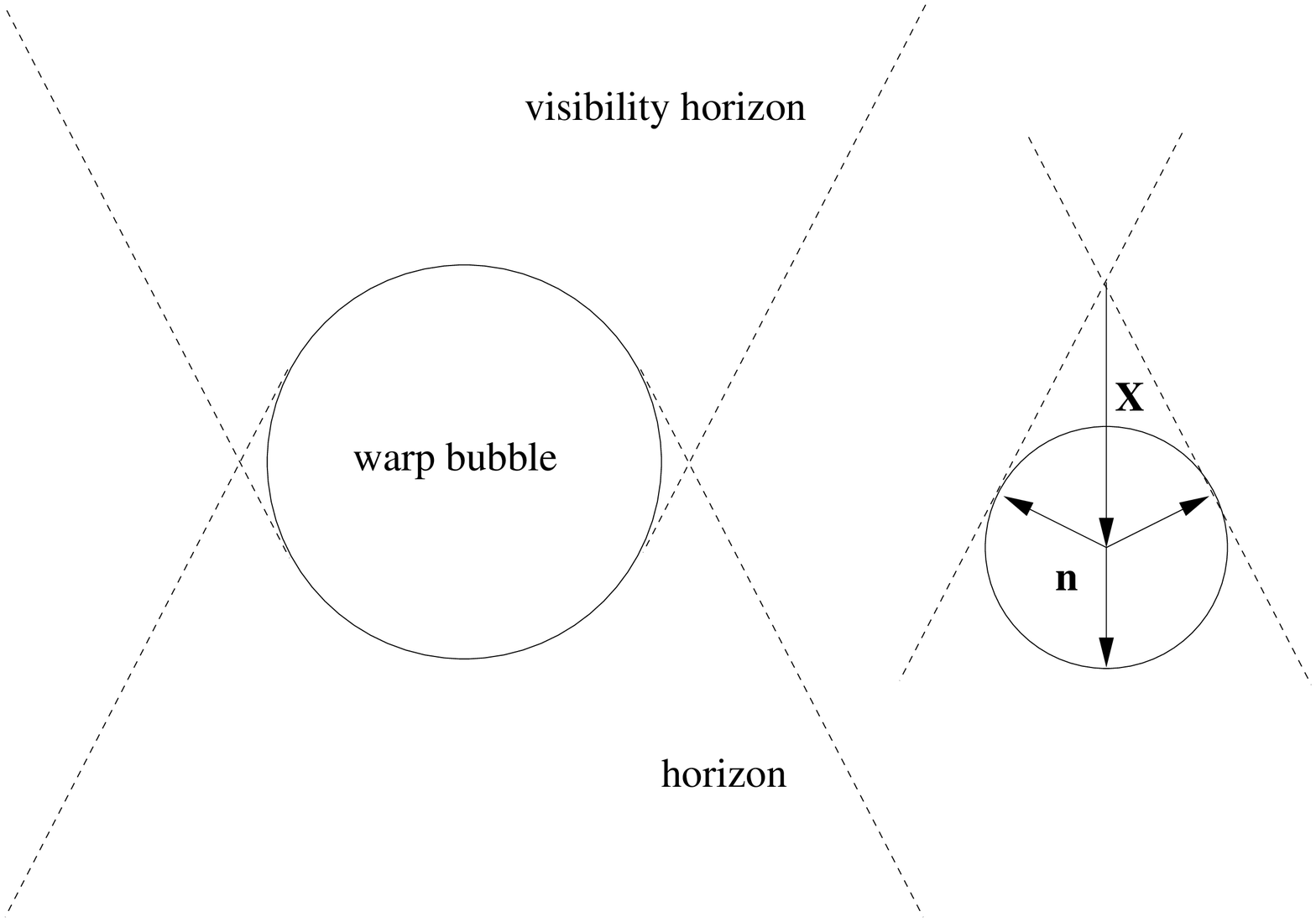}
\end{center}
\caption{Visibility horizon.}
\label{horizon3}
\end{figure}

In this approximation, it is also very easy to understand the origin of the horizon: it marks the boundary of the set of points accessible to light rays emerging from the warp bubble. Also, it is clear that the only some points which are visible from inside the warp bubble; they are delimited by a surface similar to the horizon, which we call the {\em visibility horizon} (see figure \ref{horizon3}). Notice that this observer will see nothing at all from directions beyond the direction of the visibility horizon.

We now compute redshifts. Recall that the 4-velocity of Eulerian observers satisfies $n_a = -dt$; consequently, the energy of a particle with 4-velocity
\[
U^a=\dot{t}\frac{\partial}{\partial t}+\dot{x}\frac{\partial}{\partial x}+\dot{y}\frac{\partial}{\partial y}+\dot{z}\frac{\partial}{\partial z}
\]
is just $-n_aU^a=\dot{t}$. From the geodesic equations, we see that
\[
\frac{\partial{\bf X}}{\partial t}={\bf 0} \Rightarrow \frac{d}{d\tau}\left(-\dot{t}-{\bf X} \cdot \left(\dot{\bf x}-\dot{t}{\bf X}\right) \right) =0
\] 
and hence
\[
E \left( 1+{\bf X} \cdot{\bf n} \right) = E_0
\]
where $E$ is the energy measured by the Eulerian observers and $E_0$ is a constant (corresponding to the energy measured by Eulerian observers when ${\bf X}={\bf 0}$, {\em i.e.}, inside the bubble). Thus the Eulerian observer at the centre of the warp bubble will see a blueshift of $1+v_s$ for light directly in front of him, decreasing to $1$ for light coming at $90^\circ$, and to zero for light at the edge of the visibility horizon. Conversely, light emitted by the Eulerian observer at the centre of the warp bubble towards his back will suffer a redshift of $1+v_s$; light rays emitted in directions approaching the horizon will become increasingly blueshifted, reaching infinite blueshift at the horizon.

%
%
%%%%%%%%%%%%%%%%%%%%%%%%%%%%%%%%%%%%% Bibliography &&&&&&&&&&&&&&&&&&&&&&&&&&&&&
%

\end{document}